\documentclass[prl,twocolumn,english,superscriptaddress]{revtex4-1}

\usepackage[T1]{fontenc}
\usepackage[latin9]{inputenc}
\usepackage{color}
\usepackage{babel}
\usepackage{latexsym}
\usepackage{float}
\usepackage{amsmath}
\usepackage{graphicx}
\usepackage{times}   
\usepackage{esint}
\usepackage[unicode=true,pdfusetitle,
 bookmarks=false,colorlinks=true,citecolor=blue,urlcolor=blue,linkcolor=red]{hyperref}
\usepackage{multirow}

\makeatletter

\@ifundefined{textcolor}{}
{%
 \definecolor{BLACK}{gray}{0}
 \definecolor{WHITE}{gray}{1}
 \definecolor{RED}{rgb}{1,0,0}
 \definecolor{GREEN}{rgb}{0,1,0}
 \definecolor{BLUE}{rgb}{0,0,1}
 \definecolor{CYAN}{cmyk}{1,0,0,0}
 \definecolor{MAGENTA}{cmyk}{0,1,0,0}
 \definecolor{YELLOW}{cmyk}{0,0,1,0}
}

\@ifundefined{date}{}{\date{}}
\AtBeginDocument{
  
}
\makeatother

\newcommand{\SAVE}[1]{}

\begin{document}
\renewcommand\abstractname{}

\title{Response to comment by Khomskii et al. on ``Spin--lattice coupling and the emergence of the trimerized phase in the $S=1$ kagome antiferromagnet Na$_2$Ti$_3$Cl$_8$''}
\author{Arpita Paul}
\affiliation{Department of Chemical Engineering and Materials Science,University of Minnesota, Minneapolis, Minnesota 55455, USA}
\author{Chia-Min Chung}
\affiliation{Department of Physics and Arnold Sommerfeld Center for Theoretical Physics,
Ludwig-Maximilians-Universitat Munchen, Theresienstrasse 37, 80333 Munchen, Germany}
\author{Turan Birol}
\email{tbirol@umn.edu}
\affiliation{Department of Chemical Engineering and Materials Science,University of Minnesota, Minneapolis, Minnesota 55455, USA}
\author{Hitesh J. Changlani}
\email{hchanglani@fsu.edu}
\affiliation{Department of Physics, Florida State University, Tallahassee, Florida 32306, USA}
\affiliation{National High Magnetic Field Laboratory, Tallahassee, Florida 32310, USA}
\date{\today}

\begin{abstract}
\end{abstract}

\maketitle
In Ref.~\cite{Paul2020}, using a combination of density functional theory (DFT), exact diagonalization (ED), and density matrix renormalization group (DMRG) calculations, we explained the trimerization induced polar structural transition in Na$_2$Ti$_3$Cl$_8$ as a result of spin-lattice coupling and a trimerized simplex magnetic phase that was predicted to emerge in spin--1 kagome antiferromagnets in Ref. \cite{Changlani_PRB15, vondelft_PRB15}. In Ref.~\cite{comment}, the authors state that this explanation is not justified, and instead assert 
that an orbital-driven Peierls mechanism is responsible for it. In order to support this assertion, the authors rely on the structural instabilities 
present in the nonmagnetic DFT calculations we presented in Ref.~\cite{Paul2020}. We acknowledge that while symmetry does permit the possibility of an orbital--driven Peierls mechanism as the origin of the structural transition in Na$_2$Ti$_3$Cl$_8$, we argue below that the existing experimental data and our first principles calculations strongly suggest a magnetic mechanism. 

The comment~\cite{comment} notes ``it is actually not clear how one should deal with such situations as in Na$_2$Ti$_3$Cl$_8$ -- starting from the localized limit or, vice versa''. 
This is an 
important point since the orbitally driven Peierls mechanism is most active when the system is ``close to the itinerant state, (e.g., have an insulator-metal transition)'' \cite{Khomskii2005}. The examples mentioned in Ref. \cite{comment} are either 4d transition metal systems (e.g. Zn$_2$Mo$_3$O$_8$); or host 3d transition metals but in highly connected lattices, for example triangular lattices of edge-sharing octahedra (e.g. LiVS$_2$). The compound that we study,  Na$_2$Ti$_3$Cl$_8$, is a Mott insulator at room temperature with 3d Ti cations on a kagome lattice, and there is no indication of a metal-insulator transition (or proximity to one). The material has a deep green color \cite{Kelly2019}, which suggests a large gap. Additionally, our Wannier calculations performed in the nonmagnetic state give an inter-atomic hopping no larger than 260 meV in the high temperature phase, which is significantly smaller than the Hund's coupling. These numbers justify the use of a localized orbital starting point for the high temperature phase of Na$_2$Ti$_3$Cl$_8$, as opposed to use of an itinerant picture \cite{Streltsov2016}. 

Our DFT calculations predict an unstable $\Gamma_2^-$ phonon when no $+U$ correction is employed and spin polarization is not allowed, i.e. the system is treated as nonmagnetic. The comment objects to us disregarding this result as unphysical, and states that this is ``nothing else but a manifestation of the Peierls instability'' 
present in the real material. We stand by our claim that the phonons in the magnetically ordered, insulating phase we obtain with DFT$+U$ gives 
closer results to the real material which is a Mott insulator at room temperature. It is well known in the first principles community that DFT often predicts fictitious lattice instabilities in transition metal compounds when magnetic moments are not taken into account \cite{Savrasov2003, Paul2019Rev}. The reason is that the non-spin polarized electronic state of DFT is often a bad approximation to the paramagnetic state, where there are fluctuating but well formed local moments. Even elemental iron develops lattice instabilities if magnetic moments are not taken into account carefully \cite{Kormann2012, Leonov2012}. We also note that the phonon dispersion calculated using nonmagnetic DFT (not shown) predicts multiple lattice instabilities throughout the Brillouin zone in Na$_2$Ti$_3$Cl$_8$, which strongly supports our claim that the $\Gamma_2^-$ instability in the DFT calculations is not a result of the Peierls instability, but is instead a result of the shortcomings of nonmagnetic DFT.  The comment also notes the case of VO$_2$, which is still under active investigation. 
It is metallic above its structural transition which coincides with a correlation related metal insulator transition~\cite{Zheng_Wagner}, and it can be explained properly neither by DFT nor by DFT$+U$. 

We conclude by addressing the suggestion that our explanation is less ``conceptually simple and straightforward'' than the orbital-Peierls mechanism \cite{comment}. 
%
%
We underline that ring and biquadratic exchange terms emerge from the DFT$+U$ calculations and their existence is broadly consistent with Ref.~\cite{Hoffmann2020} who recently 
provided a detailed justification for ring-exchange terms. Our ED and DMRG calculations suggest that these 
additional terms are strong enough to suppress the formation of the trimerized phase, 
which naturally arises as the ground state of the nearest neighbor spin--1 Heisenberg model~\cite{Changlani_PRB15, vondelft_PRB15}. 
%

We thank Tyrel McQueen and Juan Chamorro for several illuminating discussions. H.J.C. acknowledges support from Florida State University and the National High Magnetic Field Laboratory. The National High Magnetic Field Laboratory is supported by the National Science Foundation through NSF/DMR-1644779 and the state of Florida. H.J.C. is also supported by NSF-CAREER Grant No. DMR-2046570.
The work at the University of Minnesota was supported by NSF DMREF Grant No. DMR-1629260.

\onecolumngrid
\bibliography{response_Khomskii}

\end{document}